\documentclass[twocolumn]{jjap2}
\usepackage{txfonts}

\title{Lattice thermal conductivity in isotope diamond asymmetric superlattices}

\author{Hsu Kai Weng$^{1}$, Akira Nagakubo$^{1}$, Hideyuki Watanabe$^{2}$, and Hirotsugu Ogi$^{1}$\thanks{E-mail: ogi@prec.eng.osaka-u.ac.jp}}

\inst{
$^{1}$Graduate School of Engineering, Osaka University, Suita, Osaka 565-0871, Japan\\
$^{2}$Nanoelectronics Research Institute, National Institute of Advanced Industrial Science and Technology (AIST), Central 2, 1-1-1 Umezono, Tsukuba, Ibaraki 305-8568, Japan\\
}

\abst{
We study lattice thermal conductivity of isotope diamond superlattices consisting of $^{12}$C and $^{13}$C diamond layers at various superlattice periods.  It is found that the thermal conductivity of a superlattice is significantly deduced from that of pure diamond because of the reduction of the phonon group velocity near the folded Brillouin zone. The results show that asymmetric superlattices with different number of layers of $^{12}$C and $^{13}$C diamonds exhibit higher thermal conductivity than symmetric superlattices even with the same superlattice period, and we find that this can be explained by the trade-off between the effects of phonon specific heat and phonon group velocity. Furthermore, impurities and imperfect superlattice structures are also found to significantly reduce the thermal conductivity, suggesting that these effects can be exploited to control the thermal conductivity over a wide range.\\
\\
Accepted for publication in \textit{Japanese Journal of Applied Physics} on December 14, 2021.
}

\begin{document}
\maketitle
\section{Introduction}
In recent decades, multilayer thin films, including superlattices, have been intensively studied in various fields, including electronics\cite{Froyen,Menzel}, optics\cite{Sakaki}, magnetics\cite{Baibich}, and acoustics\cite{Bando, Fukuda}. The periodic structure in a superlattice produces unique physical properties, especially in electronics and optoelectronics. Heat conduction in superlattices has also been studied to develop materials with controllable thermal conductivity, especially for thermoelectric devices\cite{Hicks,Lin}. The thermal conductivity of diamond is extremely high, and if it becomes possible to control its thermal conductivity, it will lead to the creation of materials with a wide range of thermal conductivity. However, in order to maintain the inherent high stiffness and high strength of diamond, it is necessary to grow diamond epitaxially without mixing mismatched materials. We consider that the superlattice growth of isotopic diamond is effective to change only the thermal conductivity significantly keeping the high stiffness and strength owing to small lattice mismatch between $^{12}$C and $^{13}$C diamonds: Interfacial defects are usually introduced in a superlattice because of the lattice mismatch, and they significantly deteriorate the elasticity\cite{Nakamura1,Nakamura2,Nakamura3} and thermal conductivity\cite{Yao,Lee,Capinski} of the superlattice.  However, because the lattice mismatch at the interface between $^{12}$C and $^{13}$C diamonds is quite small ($\sim$0.01\%)\cite{Yamanaka}, the ambiguity related to the lattice mismatch will be negligible.  

Isotopically controlled diamonds demonstrate special physical properties: Anthony \textit{et al.}\cite{Anthony} reported that thermal conductivity is enhanced by about 50\% by reducing the isotope impurity ($^{13}$C) content to 0.1\% from the natural ratio of 1.1\%. (Natural carbon consists of 98.9\% $^{12}$C and 1.1\% $^{13}$C.) Watanabe \textit{et al.} used a chemical-vapor-deposition method\cite{Watanabe,Watanabe2,Costello} to produce $^{12}$C/$^{13}$C diamond superlattices and found a special electrical property in cathodoluminescence that cannot be explained by individual monolayer characteristics: The excitonic recombination in $^{13}$C diamond layer disappears despite that $^{13}$C diamond is involved. Such a specific property of $^{12}$C/$^{13}$C superlattice may appear in the phonon property like thermal conductivity. Based on the extremely high thermal conductivity of isotopically purified diamond ($\sim$3000 W/mK\cite{Anthony}), it will be possible to control the thermal conductivity of isotope diamond superlattice in a very wide range because superlattice structure can decrease the thermal conductivity \cite{Tamura}. 

We have previously studied the behavior of phonon propagation in isotopic diamond superlattices by picosecond ultrasonic spectroscopy and lattice dynamics calculations for symmetric superlattices with the same layer thickness of $^{12}$C and $^{13}$C diamonds\cite{WengProceeding,Weng2021}. Although a significant reduction in thermal conductivity was achieved, it was difficult to control the thermal conductivity more precisely in a symmetric superlattice.  In this study, we systematically study the heat conduction in isotope diamond superlattices with various lattice periods, including asymmetric structures, using a lattice thermal conductivity calculation, and discuss the controllability of heat conduction in detail.

\section{Theoretical calculation}
For theoretically investigating thermal conduction in superlattices, we perform the lattice thermal conductivity calculation following Tamura et al.\cite{Tamura}. It starts from phonon kinetic theory:
\begin{equation}
\kappa = \sum_{\lambda}\kappa_{\lambda} = \sum_\lambda{C(\omega)v_{\lambda,z}^2\tau_{\lambda}},
\end{equation}
where $\lambda$ denotes the phonon mode determined by the wavenumber ($k_x$, $k_y$, $k_z$), frequency $\omega$, and polarization. The total thermal conductivity is obtained from the summation of the contributions of different phonon modes $\kappa_{\lambda}$. The specific heat $C$($\omega$) can be calculated by the quantum harmonic oscillator and the Planck distribution via
\begin{equation}
C(\omega) = \frac{\partial}{\partial T}(\hbar \omega(N_0+1/2)) = \frac{(\hbar \omega)^2}{k_B T^2} \frac{\textrm{exp}(\hbar \omega / k_B T)}{[\textrm{exp}(\hbar \omega / k_B T)-1]^2}.
\end{equation}
The out-of-plane group velocity $v_{\lambda,z}$ can be obtained by lattice dynamics. In our lattice dynamics model\cite{Weng2021}, each atom is bonded by four first-nearest-neighbor atoms and twelve second-nearest-neighbor atoms, with bond-stretching and bond-bending stiffness. These values are determined by the reported experimental data on the phonon dispersion relationship\cite{Warren} and the measured elasticity\cite{Weng2021} via picosecond ultrasound spectroscopy\cite{Stoner,Thomsen,Matsuda,Weng2020}. The details of determining the bond stiffnesses, the lattice dynamics model\cite{Weng2021}, and the experimental system\cite{Tanigaki,Nagakubo2019,Kusakabe} are shown elsewhere. The interatomic distance is calculated by the reported lattice constant.\cite{Yamanaka} Since the differences in lattice constant and interatomic bond strength between $^{12}$C and $^{13}$C diamonds are very small ($\sim$0.01\% from the lattice constant\cite{Yamanaka} and $\sim$0.2\% from the elasticity\cite{Vogelgesang,Weng2021}, respectively), the only difference between the $^{12}$C diamond layer and the $^{13}$C diamond layer is the mass of the constituent atoms. The phonon mean-free time $\tau_{\lambda}$ is normally ambiguous, and we evaluated the thermal conductivity $\tilde{\kappa}$ normalized by $\tau_{\lambda}$, assuming that the phonon-mean free time is the same for each mode.  $\tilde{\kappa}$ then takes the form:
\begin{equation}
\tilde{\kappa} = \sum_{\lambda}\kappa_{\lambda}/\tau_{\lambda} = \sum_\lambda{C(\omega)v_{\lambda,z}^2}.
\end{equation}
Throughout this study, we use the notation ($n_{12C}$, $n_{13C}$) for representing a superlattice consisting of a $^{12}$C diamond layer with $n_{12C}$ layers and a $^{13}$C diamond layer with $n_{13C}$ layers \cite{Weng2021}, where the stacking direction is along [100] direction.

\begin{figure}[tb]
\begin{center}
\includegraphics[width=90mm]{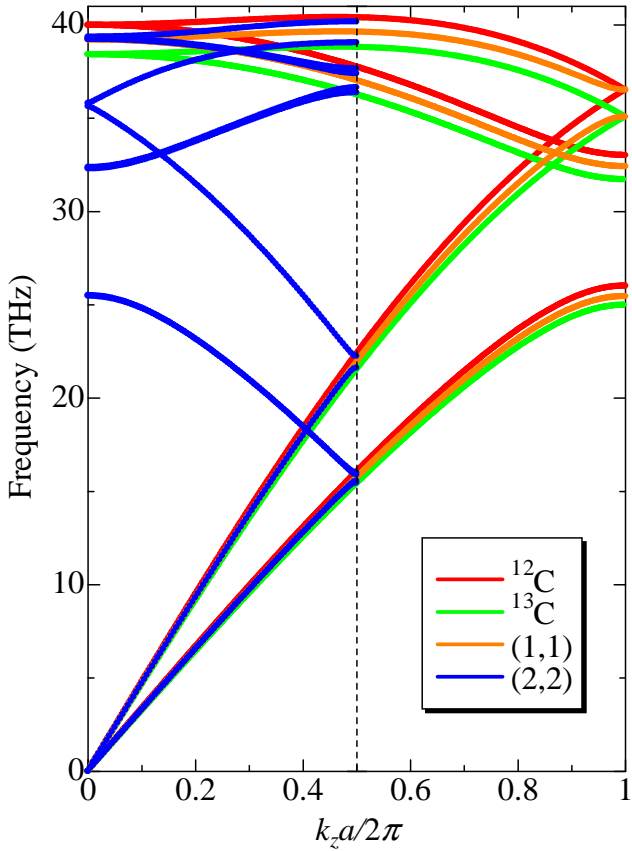}
\caption{Phonon dispersion curves calculated for $^{12}$C and $^{13}$C diamonds and for superlattices of ($n_{12C}$, $n_{13C}$)=(1,1), and (2,2). $a$ denotes the lattice constant.}
\label{DC}
\end{center}
\end{figure}

\begin{figure}[tb]
\begin{center}
\includegraphics[width=90mm]{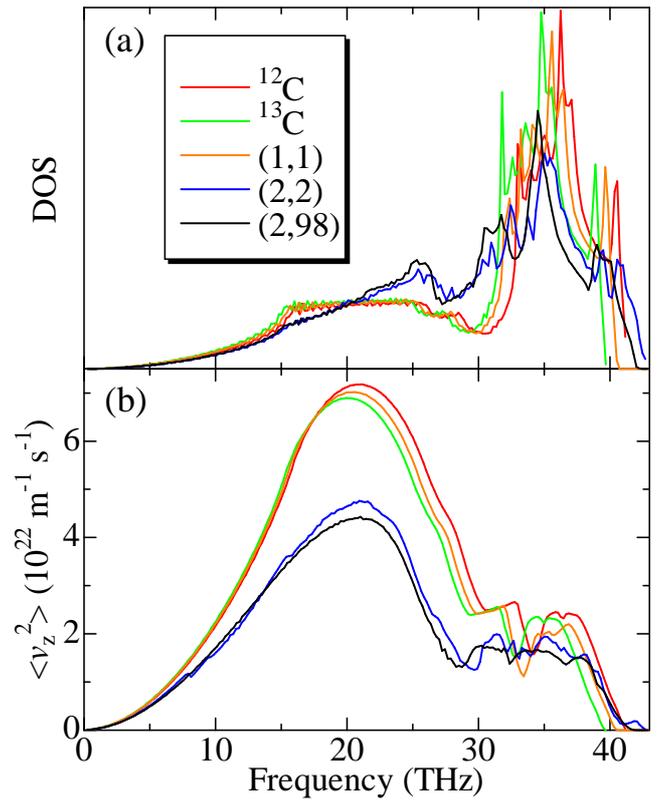}
\caption{Calculated (a) phonon DOS and (b) weighted phonon group velocity for $^{12}$C and $^{13}$C diamonds and for superlattices of ($n_{12C}$, $n_{13C}$)=(1,1), (2,2), and (2,98).}
\label{DOS}
\end{center}
\end{figure}

\section{Results and discussion}
Figure 1 presents the dispersion curves calculated for $^{12}$C and $^{13}$C diamonds, and isotope diamond superlattices of ($n_{12C}$, $n_{13C}$)=(1, 1) and ($n_{12C}$, $n_{13C}$)=(2, 2). The band gap is observed near the Brillouin zone boundary for the superlattices due to the mass difference between $^{12}$C and $^{13}$C, where the phonon group velocity is decreased. The weighted phonon group velocity, calculated by the out-of-plane phonon group velocity weighted by the phonon density of states (DOS) per unit volume as shown in Fig. 2(a), is introduced to evaluate the effect of the out-of-plane phonon group velocity as shown in Fig. 2(b). 

Figure 3 shows the temperature dependence of the normalized thermal conductivity.  The pure $^{12}$C diamond has the thermal conductivity slightly higher than that of $^{13}$C diamond because of the larger phonon group velocity due to the lighter atoms. The thermal conductivity of the isotope diamond superlattice is found to be significantly lower than that of pure $^{12}$C or $^{13}$C diamond. This reduction  is mainly explained by the decrease in the phonon group velocity near the Brillouin zone boundary. It is also found that as the superlattice period ($n_{12C}$+$n_{13C}$) increases, the normalized thermal conductivity of the superlattice decreases. However, there is an exception, that is the (1,1) superlattice, which shows the thermal conductivity as high as the value between those of $^{12}$C and $^{13}$C diamonds.  This behavior is reflected from the phonon DOS and weighted group velocity (Fig. 2) of the (1,1) superlattice, which are very different from those of other superlattices. 
In the crystal structure of diamond, the two atoms are arranged in the reference positions of the face-centered cubic unit lattice, and whether it is a $^{12}$C diamond, $^{13}$C diamond, or (1,1) superlattice, these are two atoms with the same period in the lattice structure, and their phonon dispersion relations do not change significantly (Fig. 2(a)). Therefore, the phonon group velocity does not decrease as much as in other superlattices (Fig. 2(b)).

\begin{figure}[tb]
\begin{center}
\includegraphics[width=90mm]{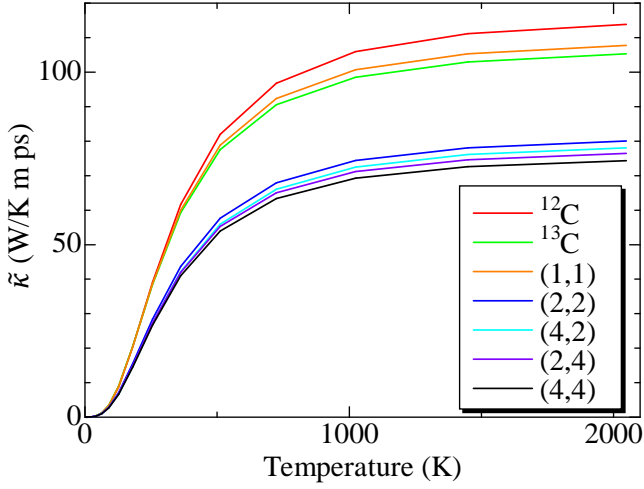}
\caption{The lattice thermal conductivity normalized by the phonon mean free time, $\tilde{\kappa}$, calculated for various isotope diamond superlattices.}
\label{Kappa}
\end{center}
\end{figure}

We then investigated the heat conduction in asymmetric superlattices. From Fig. 3, the thermal conductivity of (2,2) superlattice is higher than that of (2,4), (4,2), and (4,4) superlattices because of its shorter period, as expected. In the same period, the thermal conductivity of (2,4) superlattice is lower than that of (4,2) superlattice because of the higher $^{13}$C concentration (lower group velocity), which is also expected. However, we found that even for the same superlattice period, the thermal conductivity of the superlattice does not decrease monotonically with increasing $^{13}$C layers: Figure 4 shows the effect of the number of $^{13}$C layers ($n_{13C}$) on the thermal conductivity for three types of superlattices with the same superlattice period ($n_{12C}$+$n_{13C}$=8, 40, and 100). It should be noted that $\tilde{\kappa}$ shows a minimum. We attribute this behavior to the trade-off between the heat-capacity increase and the group-velocity decrease by increasing the $^{13}$C layer.  Lower frequency modes show higher heat capacity according to Eq (2), and, since $^{13}$C is heavier than $^{12}$C, it causes lower frequency modes and therefore results in a higher heat capacity. This effect contributes to increasing the thermal conductivity.  Figure 5 presents the phonon DOS for 100-atom period superlattices. The (2,98) superlattice has a high content of $^{13}$C and its phonon modes are close to those of pure $^{13}$C (Fig. 2(a)), showing the mode shift to lower frequencies.  Thus, its overall phonon frequencies are lower than those of (50,50) and (98,2). On the other hand, the group velocity decreases as the $^{13}$C content increases, lowering the thermal conductivity. Thus, a superlattice with a specific $^{13}$C content will show the minimum thermal conductivity under the same superlattice period.  The minimum thermal conductivity happens at the $^{13}$C content of about 0.75. The similar behavior is obtained for the other superlattice period (Fig. 4). 

For the mean free time of phonons, we have previously found that the mini-umklapp process is the main scattering mechanism in diamond superlattices\cite{Weng2021}, and we consider its effect in the following. The relaxation time for this process is given by\cite{Ren}
\begin{equation}
\tau_{U}^{-1}=B_{U}T^3\omega^2+c/L,
\end{equation}
where
\begin{equation}
B_{U}=C_{U}(e^{-\frac{\Theta}{sT}}+(\Delta M/M)^2 (n_{12C}+n_{13C})^{-2}e^{-\frac{\Theta}{s(n_{12C}+n_{13C})T}}).
\end{equation}
$c$, $L$, $\Theta$, and $\Delta M/M$ denote the speed of sound, characteristic length, Debye temperature, and the fraction of the mass difference, respectively. $C_U$ and $s$ are constants. Since these parameters are identical for superlattice diamonds, the $B_{U}$ value will be unchanged under the condition that the superlattice period $n_{12C}$+$n_{13C}$ is the same. Therefore, the trend of the normalized thermal conductivity revealed here for asymmetric superlattices is expected to appear in the actual thermal conductivity for the same superlattice period.

\begin{figure}[tb]
\begin{center}
\includegraphics[width=90mm]{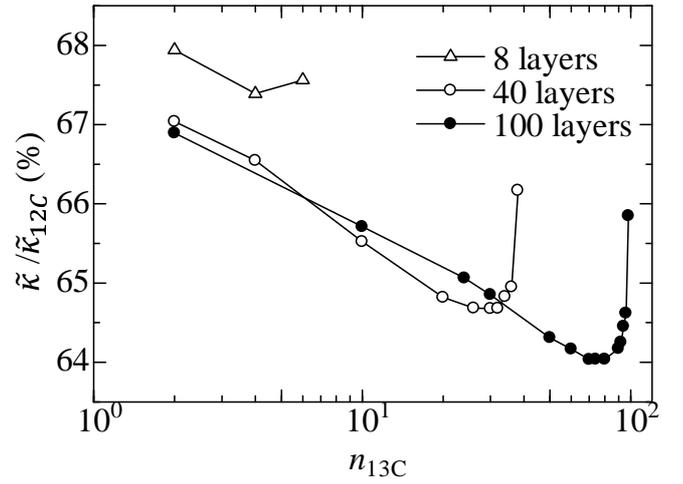}
\caption{Dependence of the lattice thermal conductivity on the number of $^{13}$C layers ($n_{13C}$) for three types of superlattices with $n_{12C}$+$n_{13C}$=8, 40, and 100 at 297 K.  $\tilde{\kappa}_{12C}$ denotes the normalized thermal conductivity for $^{12}$C diamond.}
\label{Asym}
\end{center}
\end{figure}

\begin{figure}[tb]
\begin{center}
\includegraphics[width=90mm]{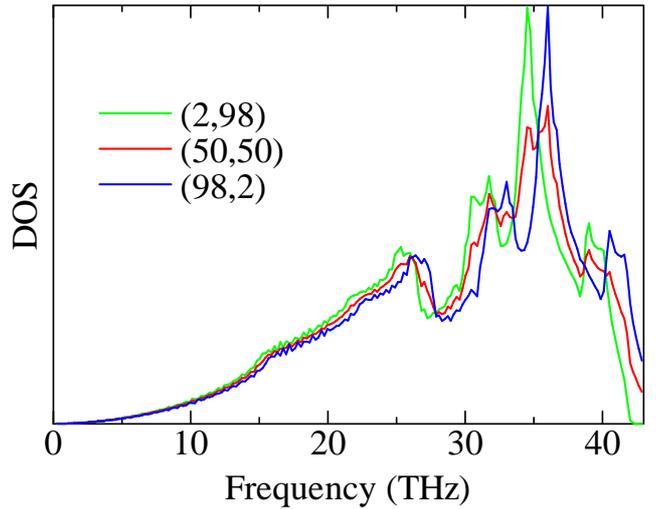}
\caption{Phonon DOS of (2,98), (50,50), and (98,2) superlattices.}
\label{DOS2}
\end{center}
\end{figure}

Next, we investigate the influence of the two $^{13}$C layers on the thermal conductivity of the superlattices with different $^{12}$C layers ($n_{12C}$, 2), and vice versa (2, $n_{13C}$). This calculation is made to investigate the apparent effect of isotopic impurities on the thermal conductivity, where the two layers can be considered as the impurity unit.  The results are in Fig. 6. The presence of the impurity layers significantly reduces the thermal conductivity, because it increases the superlattice period and therefore decreases the phonon group velocity. At the impurity content of 1\%, the thermal conductivity is decreased by $\sim$33\% from that of the pure diamond, which well agrees with the experimental observation by Anthony \textit{et al.}\cite{Anthony}, in which they reported that $\sim$1\% impurity reduced the thermal conductivity by $\sim$33\%. Therefore, our calculation may be useful for evaluating the impurity effect on the thermal conductivity.  The trend that the thermal conductivity decreases in the order  ($n_{12C}$, 2), (2, $n_{13C}$), and ($n_{12C}$=$n_{13C}$) in Fig. 6 can be explained by the trade-off between the heat-capacity increase and the group-velocity decrease by increasing the $^{13}$C layer as demonstrated in Fig. 4.

We also calculated the lattice thermal conductivity of superlattices with imperfect layer structure. Figure 7 presents the results for (2,2)-like diamond superlattices involving the thicker layer than the others. There is also a decrease in the thermal conductivity ($\sim$5\%), but this is not significant compared to the effect of isotopic impurities.  Therefore, imperfection in the superlattice period will deteriorate the thermal conductivity insignificantly.

\begin{figure}[tb]
\begin{center}
\includegraphics[width=90mm]{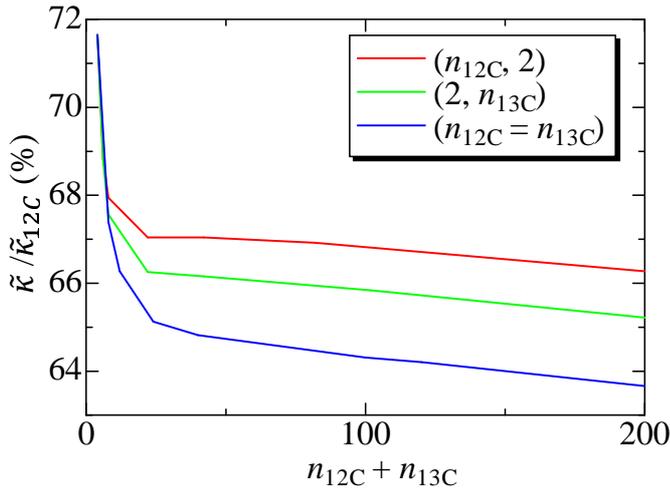}
\caption{The lattice thermal conductivity normalized by the phonon mean free time $\tilde{\kappa}$ for ($n_{12C}$, 2) and (2, $n_{13C}$), and ($n_{12C}$, $n_{13C}$) with $n_{12C}$ = $n_{13C}$ at 297 K.}
\label{N2}
\end{center}
\end{figure}

\begin{figure}[tb]
\begin{center}
\includegraphics[width=90mm]{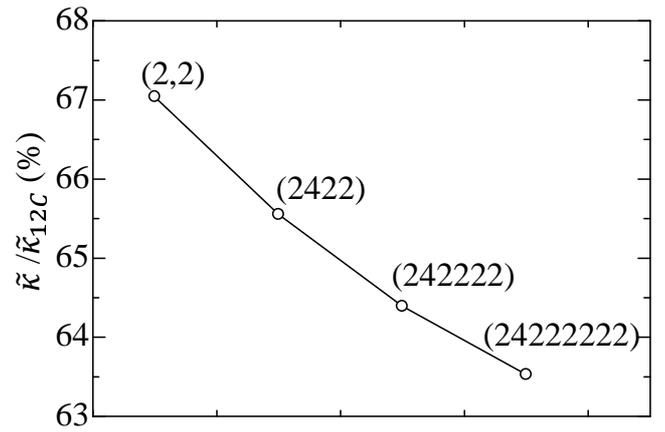}
\caption{The lattice thermal conductivity normalized by the phonon mean free time $\tilde{\kappa}$ for (2,2)-like superlattices, in which the four-atoms layer is introduced at 297 K. (2422) superlattice, for example, indicates that the numbers of layers in $^{12}$C-$^{13}$C-$^{12}$C-$^{13}$C--$^{12}$C-$^{13}$C-$^{12}$C-$^{13}$C are in the order 2-4-2-2--2-4-2-2; so on and so forth.}
\label{2422}
\end{center}
\end{figure}

\section{Conclusion}
We studied the lattice thermal conductivity for isotope diamond superlattices, consisting of $^{12}$C and $^{13}$C diamond layers with various superlattice periods, including asymmetric superlattices.  Although the calculation does not consider the phonon mean free time, the trend given in this study will be applicable to actual superlattices because the phonon scattering effect is expected to be identical at the interface. The superlattice period has a significant effect on the thermal conductivity. Because of the reduced phonon group velocity near the folded Brillouin zone, the longer the superlattice period is, the lower the thermal conductivity. Our calculation revealed that, under the same superlattice period condition, the thermal conductivity of diamond superlattice can be controllable with the asymmetric superlattice: When the ratio of $^{12}$C to $^{13}$C is $\sim$1:3, the thermal conductivity becomes the lowest value. This behavior was explained by the trade-off between the heat-capacity increase and the group-velocity decrease by increasing the $^{13}$C layer.  We also found that the isotope impurity effect can be also evaluated with our superlattice model, which resulted in the isotope impurity effect close to the experimental observation.  Similarly, when one layer becomes thicker than the other, the thermal conductivity decreases. Therefore, the imperfection in the superlattice structure and isotope impurities significantly affect the phonon group velocity and further the thermal conductivity. These effects need to be taken into account in theoretical calculations and thermal conductivity measurements.

\section*{Acknowledgment}
This study was supported by JSPS KAKENHI Grant No. JP19H00862.


\begin{thebibliography}{99}
\bibitem{Froyen} S. Froyen, D. M. Wood, and A. Zunger: Phys. Rev. B \textbf{37} (1988) 6893.
\bibitem{Menzel} D. Menzel, W. Koschinski, K. Dettmer, and J. Schoenes: Thin Solid Films \textbf{342} (1999) 312.
\bibitem{Sakaki} H. Sakaki, T. Noda, K. Hirakawa, M. Tanaka, and T. Matsusue: Appl. Phys. Lett. \textbf{51} (1987) 1934.
\bibitem{Baibich} M. N. Baibich, J. M. Broto, A. Fert, F. Nguyen Van Dau, F. Petroff, P. Etienne, G. Creuzet, A. Friederich, and J. Chazelas: Phys. Rev. Lett. \textbf{61} (1988) 2472.
\bibitem{Bando} Y. Bando, M. Toyoda, and S. Saito: J. Phys. Soc. Jpn \textbf{88} (2019) 034603.
\bibitem{Fukuda} H. Fukuda, A. Nagakubo, and H. Ogi: Jpn. J. Appl. Phys. \textbf{60} (2019) SDDA05.
\bibitem{Hicks} L. D. Hicks, T. C. Harman, and M. S. Dresselhaus: Appl. Phys. Lett. \textbf{63} (1993) 3230.
\bibitem{Lin} P. J. Lin-Chung and T. L. Reinecke: Phys. Rev. B \textbf{51} (1995) 13244.
\bibitem{Nakamura1} N. Nakamura, H. Ogi, T. Yasui, M. Fujii, and M. Hirao: Phys. Rev. Lett. \textbf{99} (2007) 035502.
\bibitem{Nakamura2} N. Nakamura, A. Uranishi, M. Wakita, H. Ogi, and M. Hirao: Jpn. J. Appl. Phys. \textbf{49}, (2010) 07HB04.
\bibitem{Nakamura3} N. Nakamura, R. Yokomura, N. Takeuchi, D. Yamakado, and H. Ogi: Jpn. J. Appl. Phys. \textbf{58}, (2019) 075504.
\bibitem{Yao} T. Yao: Appl. Phys. Lett. \textbf{51} (1987) 1798.
\bibitem{Lee} S. M. Lee, D. G. Cahill, and R. Venkatasubramanian: Appl. Phys. Lett. \textbf{70} (1997) 2957.
\bibitem{Capinski} W. S. Capinski, H. J. Maris, T. Ruf, M. Cardona, K. Ploog, and D. S. Katzer: Phys. Rev. B \textbf{59} (1999) 8105.
\bibitem{Yamanaka} T. Yamanaka, S. Morimoto, and H. Kanda: Phys. Rev. B \textbf{49} (1994) 9341.
\bibitem{Anthony} T. R. Anthony, W. F. Banholzer, J. F. Fleischer, L. Wei, P. K. Kuo, R. L. Thomas, and R. W. Pryor, Phys. Rev. B \textbf{42}, 1104 (1990).
\bibitem{Watanabe} H. Watanabe, C. E. Nebel, and S. Shikata, Science \textbf{324}, 1425 (2009).
\bibitem{Watanabe2} H. Watanabe, T. Koretsune, S. Nakashima, S. Saito, and S. Shikata, Phys. Rev. B \textbf{88}, 205420 (2013).
\bibitem{Costello} M. C. Costello, D. A. Tossell, D. M. Reece, and C. J. Brieeley, Diamond and Related Materials \textbf{3}, 1137 (1994).
\bibitem{Tamura} S. Tamura, Y. Tanaka, and H. J. Maris, Phys. Rev. B \textbf{60}, 2627 (1999).
\bibitem{WengProceeding} H. K. Weng, A. Nagakubo, H. Ogi, and H. Watanabe: Proc. 42th Symp. Ultrason. Electron. \textbf{42}, 1Pb1-3 (2021).
\bibitem{Weng2021} H. K. Weng, A. Nagakubo, H. Ogi, and H. Watanabe: Phys. Rev. B \textbf{104} (2021) 054112.
\bibitem{Warren} J. L. Warren, J. L. Yarnell, G. Dolling, and R. A. Cowley: Phys. Rev. \textbf{158} (1967) 805.
\bibitem{Stoner} R. J. Stoner and H. J. Maris: Phys. Rev. B \textbf{48} (1993) 16373.
\bibitem{Thomsen} C. Thomsen, H. T. Grahn, H. J. Maris, and J. Tauc: Phys. Rev. B \textbf{34} (1986) 4129.
\bibitem{Matsuda} O. Matsuda, O. B. Wright, D. H. Hurley, V. E. Gusev, and K. Shimizu: Phys. Rev. Lett. \textbf{93} (2004) 095501.
\bibitem{Weng2020} H. K. Weng, A. Nagakubo, H. Watanabe, and H. Ogi: Jpn. J. Appl. Phys. \textbf{59} (2020) SKKA04.
\bibitem{Tanigaki} K. Tanigaki, H. Ogi, H. Sumiya, K. Kusakabe, N. Nakamura, M. Hirao, and H. Ledbetter: Nat. Commun., \textbf{4} (2013) 2343.
\bibitem{Nagakubo2019} A. Nagakubo, S. Tsuboi, Y. Kabe, S. Matsuda, A. Koreeda, Y. Fujii, and H. Ogi, Appl. Phys. Lett. \textbf{114}, 251905 (2019).
\bibitem{Kusakabe} K. Kusakabe, A. Wake, A. Nagakubo, K. Murashima, M. Murakami, K. Adachi, and H. Ogi: Phys. Rev. Mater. \textbf{4} (2020), 043603.
\bibitem{Vogelgesang} R. Vogelgesang, A. K. Ramdas, S. Rodriguez, M. Grimsditch, and T. R. Anthony: Phys. Rev. B \textbf{54} (1996) 3989.
\bibitem{Ren} S. Y. Ren and J. D. Dow: Phys. Rev B \textbf{25} (1982) 3750.

\end{thebibliography}
\end{document}